\documentclass[9pt,twocolumn,twoside]{pnas-new}
\usepackage{graphics,color,graphicx,amsmath}
\usepackage{xr}
\usepackage{rotating}[counterclockwise]
\templatetype{pnasresearcharticle} 

\title{Information consumption and size in firms}
\author[a]{Edward D. Lee}
\author[b]{Alan P. Kwan}
\author[a]{Rudolf Hanel}
\author[c]{Anjali Bhatt}
\author[a]{Frank Neffke}
\affil[a]{Complexity Science Hub Vienna, Vienna, Austria}
\affil[b]{Hong Kong University, Hong Kong, China}
\affil[c]{Harvard Business School, Boston, USA}
\dates{This manuscript was compiled on \today}

\leadauthor{Lee} 

\significancestatement{Organisms exchange and use information to function. Social organizations like firms do too, but they are hard to measure quantitatively. We use a database with unprecedented resolution and breadth to analyze how firms consume information through reading of online news. We find that aspects of reading reveal telltale signs of organizational structure from an economy of scale (information is cheaper the more firms read), that large firms face a coordination problem because teams are redundant and repetitive readers, to that firms hold onto previous interests (larger firms have cumulatively more interests). The trends relate information consumption and firm size to establish a baseline for characterizing firms with information targets and raise questions about analogous biological structure.}

\authorcontributions{E.D.L.~designed the project. E.D.L., A.P.K., and A.B.~participated in initial discussions. E.D.L., F.H, and R.H.~discussed the results. A.P.K.~and E.D.L.~processed the data. E.D.L.~did the analysis, wrote the code, and initially drafted the manuscript. E.D.L., F.N., A.P.K., and R.H.~edited the manuscript.}
\authordeclaration{The authors declare no other competing interests.}
\correspondingauthor{\textsuperscript{2}To whom correspondence should be addressed. E-mail: edlee@csh.ac.at}


\begin{abstract}
Social and biological collectives need to exchange information to persist and to function. This happens across internal networks, whose structure represents static channels through which information flows. Less studied is the quantity and variety of information transmitted. We characterize a part of the information flow, the information going into organizations, primarily business firms. We measure what firms read using a data set of hundreds of millions of records of news articles accessed by employees across millions of firms. We measure and relate quantitatively three essential aspects: reading volume, reading variety, and firm size. First we compare volume with firm size, showing that firms grow sublinearly with the volume of their reading. The scaling means that inequality in information volume exaggerates the classic Zipf's law inequality in firm size, pointing to an economy of scale in information consumption. Then, by connecting variety and volume, we show that the firms vary in their reading habits to a limited degree. Firms above a certain size become repetitive readers, consistent with the sudden onset of a coordination cost between teams, not individual employees. Finally, we relate information variety to size to show that large firms tend to increase investments in existing areas of interest instead of divesting from them to move to new areas. We argue that this reflects structural constraints in growth. The results indicate how information consumption reflects the role of internal structure, beyond individual employees, analogous to  information processing in other social and biological systems.
\end{abstract}

\begin{document}
\maketitle

\thispagestyle{firststyle}
\ifthenelse{\boolean{shortarticle}}{\ifthenelse{\boolean{singlecolumn}}{\abscontentformatted}{\abscontent}}{}

\dropcap{I}nformation exchange facilitates collective behavior including coordination in schools of fish or flocks of birds \cite{couzinEffectiveLeadership2005,balleriniInteractionRuling2008,rosenthalRevealingHidden2015,hartnettHeterogeneousPreference2016}, conflict levels in primate society \cite{brushFamilyAlgorithms2013}, team performance \cite{wuchty2007increasing}, and organizational function \cite{nonakaKnowledgecreatingCompany1995,bhattLanguageBasedMethod2022}. In each of these cases, the transmission of information is structured by cognitive constraints \cite{simonInformationProcessing1979}, social strategy and relationships \cite{katzInvestigatingNot1982}, or simply distance \cite{katzInferringStructure2011}.
One particularly compelling example is the business firm. Individuals throughout the firm hierarchy acquire and process information to make decisions and retransmit information to different parts of the organization \cite{simonNewDevelopments1962,cohenAbsorptiveCapacity1990}. Importantly, social structure such as bridges or weak ties \cite{granovetterStrengthWeak1973,burtStructuralHoles2004} mediate the flow of information, a perspective that has inspired the statistical physics \cite{strogatzExploringComplex2001,albertStatisticalMechanics2002} and computational social science \cite{wattsCollectiveDynamics1998} of social networks. Pioneering work has gone into characterizing the structural properties of the networks. As one mapping in the study of collective knowledge, the properties of nodes include tacit knowledge, worker skills, and social skills \cite{deming2017growing}. Edges represent coworker complementarities \cite{neffke2019value}. Structural properties may reflect organization and hierarchy \cite{girvanCommunityStructure2002}. On the other end, others focus on the output of the computation done by the networks such as how patents reflect combinatorial innovation \cite{hallRecentResearch2012,jaffeMeaningPatent2000,younInventionCombinatorial2015}. In a related vein, the management literature has investigated how firms absorb \cite{cohenAbsorptiveCapacity1990}, process, and use information such through investment on R\&D and internal communications \cite{bhattLanguageBasedMethod2022}. To connect internal structure to a firm's output, we need to determine the dynamics of information flow. Yet, an obstacle specific to the business firm is that the information flow and its contents are largely unseen. Here, we are able to peer into one part of it, news consumption. By relating scaling trends in the data to one another, we take a step towards quantitatively modeling the volume and variety of information in organizations and establish quantitative measures that can help connect them to principles that mediate or shape collective behavior more broadly.

We look into the prodigious information consumption of millions of organizations globally, the vast majority of which are business firms. To do so, we analyze an extensive data set of ``intent data'' aimed at gauging customer interest \cite{kwanDoesInternet2019,kwanInstitutionalInvestor2022}. The data consists of hundreds of millions of records of content accessed by firm employees within a large universe of publishers including The Wall Street Journal, Bloomberg, Forbes, Business Insider, and CBSi, along with more specialized groups of sites such as 1105Media, ITCentral Station, and Questex. Most are anonymous but span technology, marketing, legal, biotech, manufacturing, and a wide range of business services \cite{kwanDoesInternet2019}. We focus on a two-week period between the dates of June 10 and June 23, 2018, which we expect is generally representative of the data set as we detail further in Appendix~\ref{si sec:data} in the Supplementary Information. The limited time window also ensures that proprietary preprocessing steps used to generate the data remain consistent. In principle, the data would allow us to determine which news article an anonymized employee at a firm accessed and when. For each article, we have up to ten associated topics that have been identified with a proprietary topic modeling algorithm (more details in reference \citenum{kwanDoesInternet2019}). These different properties permit us to analyze employee reading at different scales of resolution from the individual articles, which then belong to sets of content pages or ``sources,'' and that may overlap in broader ``topics'' as diagrammed in Figure~\ref{gr:scales} in the Supplementary Information \cite{hobergConglomerateIndustry2018, hobergTextBasedNetwork2016}. Importantly, the comprehensive nature of the data set allows us to obtain a multiscale portrait of how firms seek out information down to the individual acts of information acquisition.

We focus on large-scale, population-wide aspects and establish trends between the volume of information consumption, its variety, and measures of firm size. In the first part of the paper, we relate reading volume with firm size. We find that firm size scales sublinearly with volume, leading to an inequality of information consumption that exaggerates the Zipf's law inequality in firm size \cite{axtellZipfDistribution2001}. This suggests that reading volume reflects a different organizational structure than that reflected in the usual size metrics. Then, we relate volume with variety to find that large firms tend to be redundant readers. We relate the limited variety of reading to coordination limits in firms, and we propose that the scaling of teams of employees is a crucial part of predicting how diversely large firms read. Finally, by exploring the relation between variety and firm size, we predict two qualitative extremes of firm reading strategies, where either firms'  asset-to-topic ratio increases or the ratio decreases. We predict that nearly all firms tend to the former. Generally, the quantitative model aligns with conventional wisdom for how firms focus on core competencies \cite{teeceUnderstandingCorporate1994,hidalgoEconomicComplexity2021}. Interestingly, the increasing concentration of assets per topic is in contrast with the economy of scale for reading volume, or elementary bits of information, suggesting that the way that firms use information at the scale of employee is fundamentally different from its use at the scale of the organization. This echoes our main finding that the use of information consumption reflects properties of the firm as an information processing machine that leaves quantitative traces of organizational structure.

\section*{Economy of scale in information}
\begin{figure}
	\includegraphics[width=\linewidth]{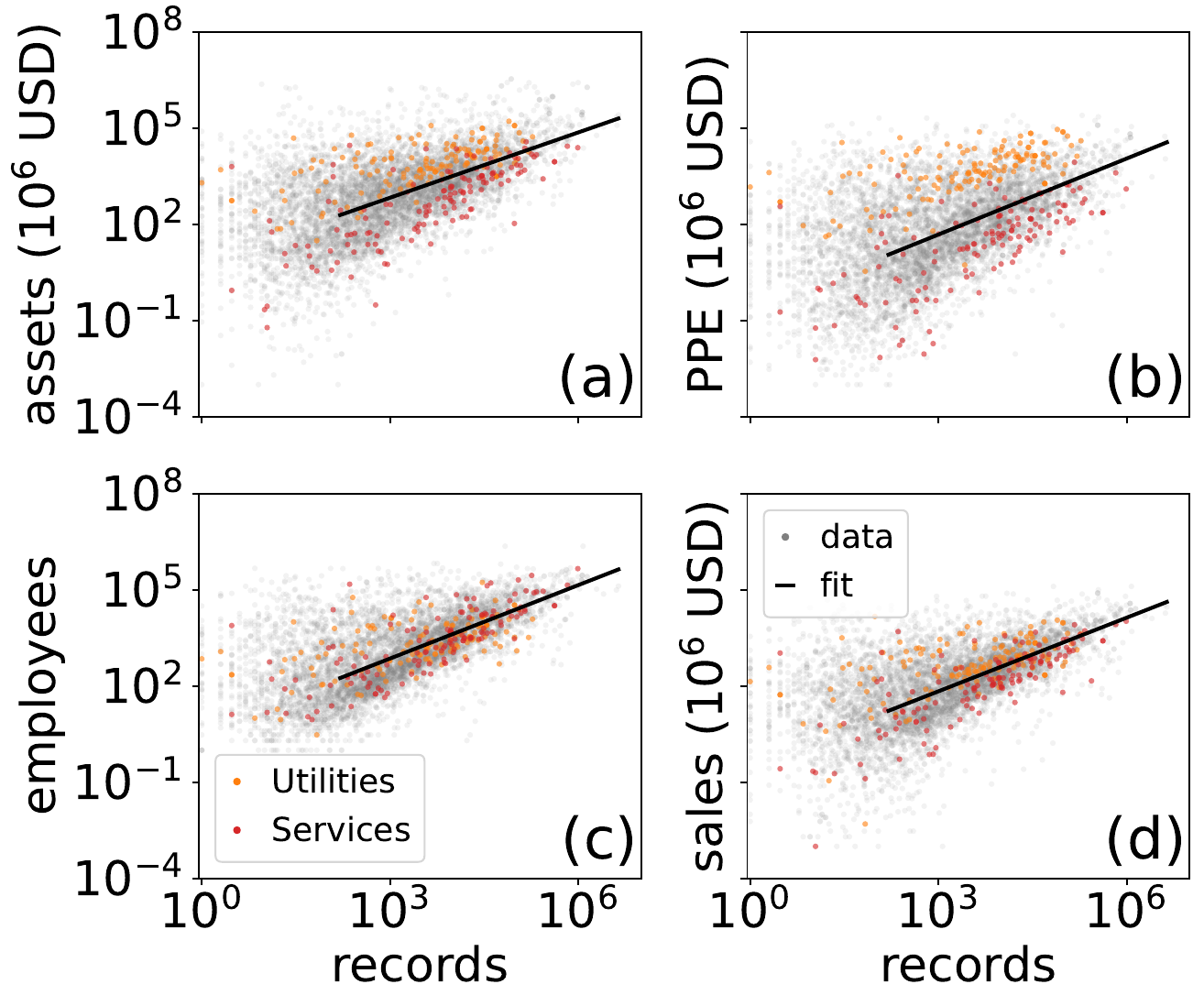}
	\caption{Scaling of firm size measures (a) assets, (b) plants, property, equipment (PPE), (c) employees, and (d) sales with record count. NAICS sectors Utilities (orange), Professional, Scientific, and Technical Services (red), and all other (gray) firms. Black line shows a power law fit to Eq~\ref{eq:scaling} with exponents (a) $\beta=0.68\pm0.02$, (b) $\beta=0.79\pm0.02$, (c) $\beta=0.76\pm0.01$, and (d) $\beta=0.77\pm0.01$ using one standard deviation from bootstrapped fits as error bars. Fitting range $R\geq160$ given by the fit to the distribution in Figure~\ref{gr:power laws}e.}\label{gr:capital}
\end{figure}

As a first step to characterizing how firms seek out information, we consider the relationship between the volume of information acquisition, the number of times a firm is recorded in the database, or ``records,'' and conventional measures of firm capital. In Figure~\ref{gr:capital}, we plot for public firms in the COMPUSTAT database the value of firm assets, PPE (plants, property, equipment), the number of employees, and sales against the number of records. Building on previous work on scaling in firms \cite{zhangScalingLaws2021}, we consider a power law relationship between economic measure $Y$ and records $R$,
\begin{align}
	Y = AR^\beta,\label{eq:scaling}
\end{align}
for some positive constant $A$ and positive exponent $\beta$. This is equivalent to a regression on a logarithmic scale, where the slope is $\beta$ and the $y$-intercept $\log A$ when Eq~\ref{eq:scaling} is transformed to $\log Y = \beta \log R + \log A$. Importantly, the exponent $\beta$ is independent of the units of $X$ and $Y$, the conversion between which is separately captured in $A$. We find that all economic measures scale sublinearly with the number of records with scaling exponents of about $\beta\approx3/4$ except for assets which scale somewhat slower as $\beta=0.68\pm0.02$ (Table~\ref{si tab:exponents}). Furthermore, it tends to be the case that a category of service firms (as indicated by North American Industrial Classification System codes, or NAICS, red points falling below the fit line) have more records per unit of economic measure as compared to firms in Utilities (orange points falling above the fit line). They are not as clearly differentiated employees, but the estimate of the number of employees in COMPUSTAT is known to be poor. While the services category includes a broad swathe of industries and thus displays wide variation along the vertical, the trend is consistent with the notion that firms in the service sector are on the whole more knowledge-intensive compared to utilities. 

\begin{figure*}\centering
	\includegraphics[width=.9\linewidth]{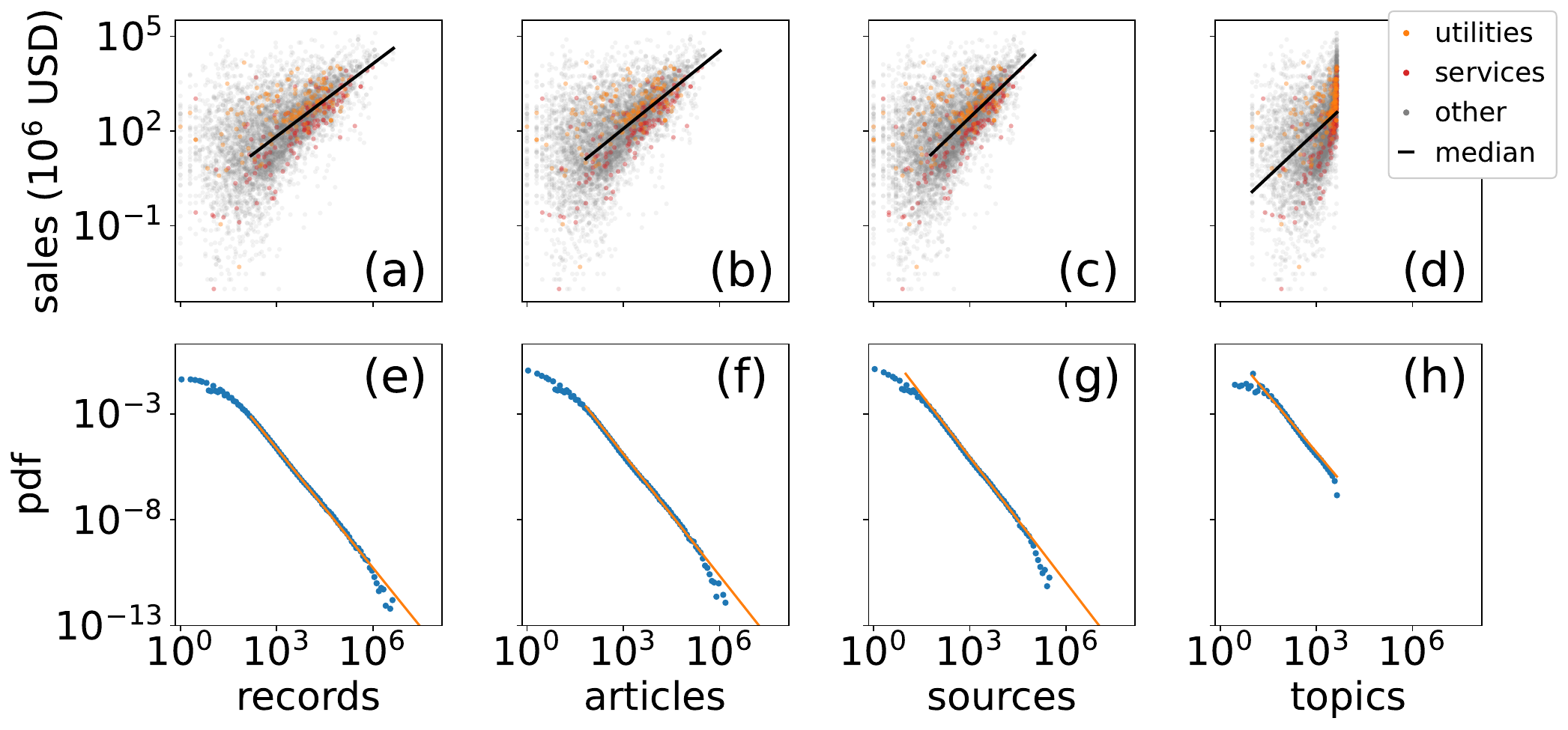}
	\caption{(a-d) Scaling of sales with information variables. Mining (green), service (red), and all other (gray) firms. Black line is the scaling fit. Scaling exponents are $\beta=0.77\pm0.01$, $\beta=0.82\pm0.02$, $\beta=0.97\pm0.02$, and $\beta=0.95\pm0.03$ (not shown) using bootstrapped error bars. Fits are only to data points above the lower cutoff given by the fit to the distributions in the panel directly below except for topics, where the lower cutoff can be set {\it a priori} because the topic relevancy vector is of size 10. (e-h) Distribution of the number of records $\alpha=1.88$, articles $\alpha=1.92$, sources $\alpha=1.97$, and topics $\alpha=1.80$ per firm shows power law scaling in the tails. A standard fitting procedure involving maximum likelihood for the exponent $\alpha$ with the Kolmogorov-Smirnov statistic for the lower bounds returns $x_{\rm min}=160$, $x_{\rm min}=69$, and $x_{\rm min}=61$ \cite{clausetPowerLawDistributions2009}. For topics, the lower bound is fixed at $x_{\rm min}=10$, and our simplified scaling model does not capture the curvature in panel d. See Table~\ref{si tab:exponents} for further exponents.}\label{gr:power laws}
\end{figure*}

A record, however, is a rather rudimentary measure of information flow. Within any given record, we are also given the article that was accessed, the page on which it published or ``source,'' and the topics that were relevant to the article, the latter a derivative measure obtained from proprietary topic modeling (see Appendix~\ref{si sec:data} for more details). Since these constitute progressively larger groupings of individual records, we might anticipate the economic size to scale differently with the information measures at coarser granularity. As we show in Figure~\ref{gr:power laws}, we find that sales grow faster with the number of articles ($\beta=0.82\pm0.02$), with sources ($\beta=0.97\pm0.02$), and with topics (${\beta=0.95\pm0.03}$), respectively, using a least-squares fit on logarithmic axes. The sublinearity with records indicates that the typical increase in elementary counts of information access grows faster than the increase in sales such that the effective cost of accessing {\it new} information decreases with reading volume or articles, but this is not the case for sources and topics.

Importantly, the scaling between economic and information measures is consistent with the heavy-tailed distributions of information search. This would mean that our results extracted for a small subset of public firms in COMPUSTAT are consistent with the distribution from the millions of firms in the reading data. We check this by using the observation that the distribution of firm receipts, here given by sales $Y$, follows a Zipf's law such that ${p(Y) \sim Y^{-2}}$ \cite{axtellZipfDistribution2001}. Then, if it is the case that the distribution of information quantity $X$ obeys ${p(X) \sim X^{-\alpha}}$, we can use the transformation $p(S)dS = p(X)dX$ and the scaling relation $Y=AR^\beta$ to obtain the exponent relation
 \begin{align}
 	\alpha &= \beta + 1.\label{eq:exp relation}
\end{align}
Using the values of $\beta$ found above, we obtain the predictions $\hat\alpha=1.77\pm0.01$, $\hat\alpha=1.82\pm 0.02$, $\hat\alpha=1.97\pm0.02$, and $\hat\alpha=1.95\pm0.03$.

Next, we fit the probability distributions by the number of records, articles, and topics in Figure~\ref{gr:power laws} using a standard method \cite{clausetPowerLawDistributions2009}, which gives us a direct estimate of $\alpha$ instead of from using the right hand side of Eq~\ref{eq:exp relation}. From this method, we obtain $\alpha=1.88\pm0.00$, $\alpha=1.92\pm0.00$, $\alpha=1.97\pm0.01$, and $\alpha=1.80\pm0.00$ for records, articles, sources, topics, respectively. We note that the fit to topics provides a negative check because it should fail. After all, the fit to topics is statistically distinguishable from a power law, is limited in its range, and the estimate of $\beta$ from panel d does not account for the curvature in the median sales by topic. Accordingly, we find that the exponent scaling relation in Eq~\ref{eq:exp relation} is least well satisfied for topics with error of $\Delta \equiv \hat\alpha-\alpha = 0.15\pm 0.03$, but is better satisfied for the remaining quantities $\Delta =-0.11\pm0.01$, $\Delta =-0.10\pm0.02$, and $\Delta =0.00\pm0.02$ for records, articles, and sources, respectively. While one has to be careful with startups and the smallest firms which show much more variability in information and economic behavior \cite{zhangScalingLaws2021}, our scaling models establish a basis for comparison relative to an expected trend. Indeed, that the exponents are close to satisfying the exponent relation Eq~\ref{eq:exp relation} --- despite $\beta$ having been extracted for a small subset of public firms in COMPUSTAT and $\alpha$ from all firms in the reading data --- indicates that our scaling approximation is reasonably self-consistent. 

The exponents $\alpha<2$ for records and articles indicates heavy-tailed distributions of firm information search and thus strong inequality in how firms read. In contrast, firm size distributions have an exponent $\alpha=2$. This means that along elementary measures of information, but not the aggregated measures of sources or topics, we find that the largest firms have a disproportionately larger information footprint compared to the disparity in economic size.

\section*{Limits to information variety}

\begin{figure}[t]\centering
	\includegraphics[width=\linewidth]{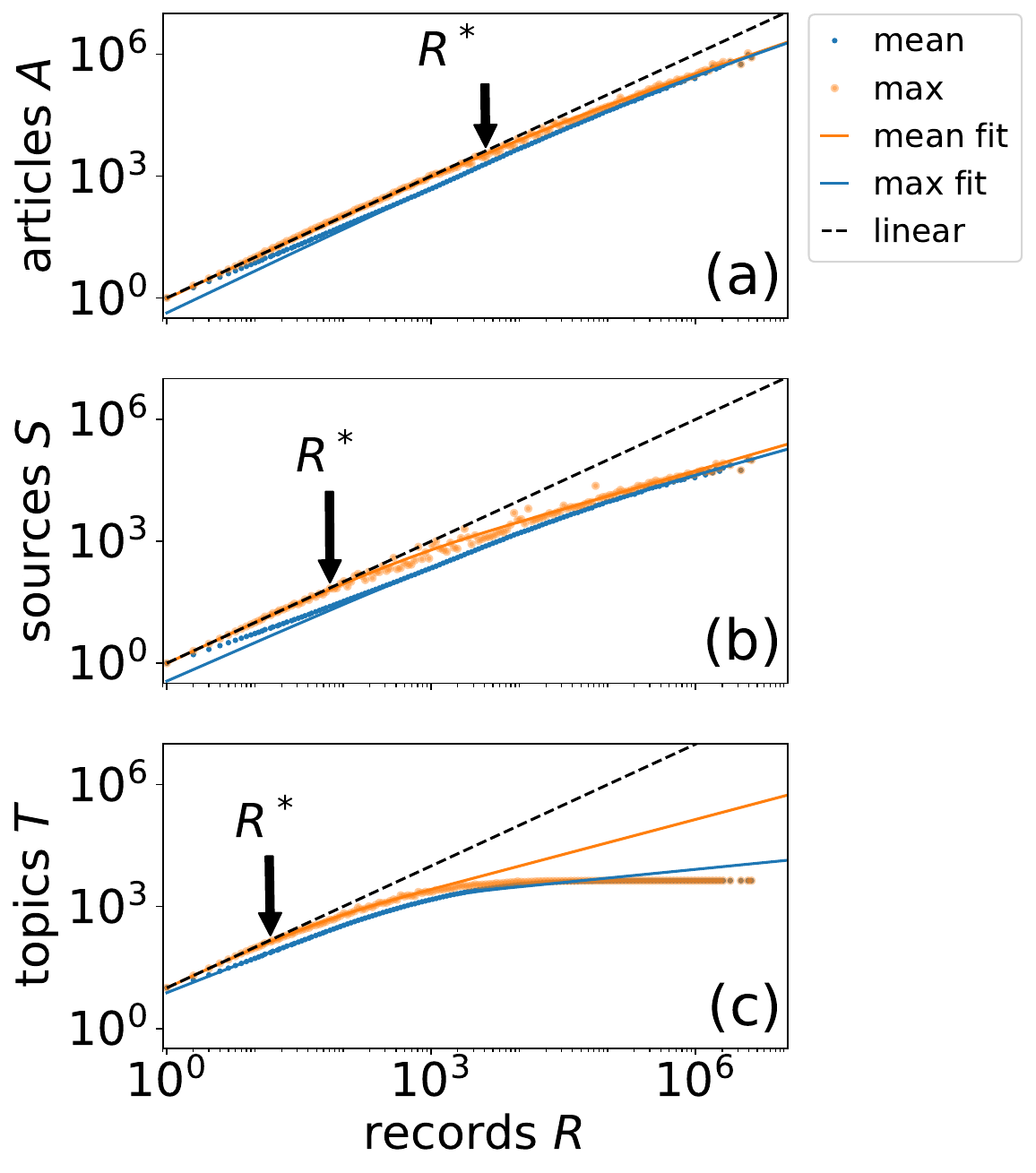}
	\caption{Diversity of information collected by firms as Heaps' plots for firm reading. Growth of (a) articles $A$, (b) sources $S$, and (c) topics $T$ with number of records $R$ along with fits of information-overlap model as lines. For a given logarithmic bin $R$, we show the firms with maximal variety in reading (orange markers) and firms with mean reading variety (blue markers). While the total number of articles ($A\sim40$ million) and sources ($S\sim800{,}000$) is much greater than what any single firm accesses in the subsample, the number of topics is bounded at $T=4{,}338$. To avoid overfitting to the cutoff, we scan over a range of values and only take fit values that are of sufficiently low error and do not violate physical limits (more details in Appendix~\ref{si sec:heaps}). Estimated team size scaling exponents are $b\approx 3/10$ for both mean and max curves for articles. For sources, $b\approx1/3$ and $b\approx2/5$ for mean and max, respectively. For topics, $b\approx1/2$ and $b\approx2/3$. Points $R^*$ at which the maximal curves fall below 90\% of the 1:1 line are indicated with arrows $R=1{,}645$ records for articles, $R=41$ records for sources, and $R=18$ records for topics.}\label{gr:heaps}
\end{figure}

More reading does not necessarily imply new information for the firm. As we show in the Heaps' plot in Figure~\ref{gr:heaps}, we compare the number of unique articles $A$, sources $S$, or topics $T$ with the total number of times the firm accessed content, or left a record $R$ \cite{heapsInformationRetrieval1978}. For a given logarithmic bin for records $R$, we plot the average number of items read by firms (blue markers) below the maximum (orange). Both data points grow sublinearly with $R$ in every panel. This is sensible because we know that a single employee might read the same article several times or share it with colleagues such that this plot would at maximum trace the black, dashed, one-to-one line. Furthermore, the three Heaps' plots all show qualitatively similar patterns of two distinct regimes: the least read (small $R$) firms saturate the variety of articles and the most read (large $R$) firms fail to saturate this curve. This means that small firms with the most diverse reading tendencies saturate the maximum number of articles they could read per record but that at a larger size, the most diverse readers peel away from linear growth, indicating that the same articles $A$, sources $S$, and topics $T$ are reread.

\begin{figure}[t]
	\includegraphics[width=\linewidth]{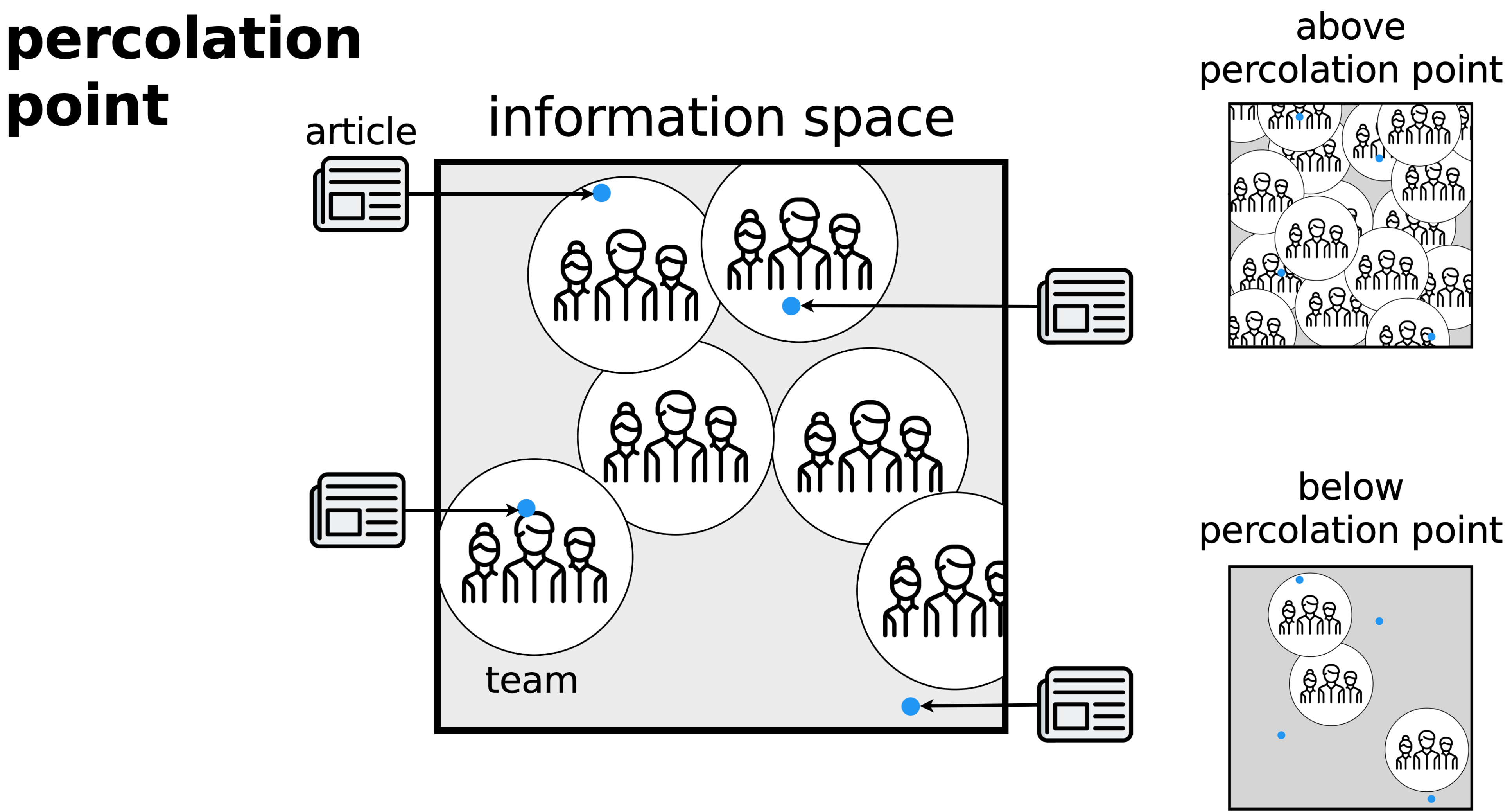}
	\caption{Diagram of information processing model. Firm consists of organizational units, each with a range of expertise in a high-dimensional space of information, here projected onto two dimensions. If an article falls into the range of expertise of a unit, the firm can realize a benefit (Eq~\ref{eq:info benefit}). In small firms below the percolation point, teams hardly fill the space, and in large firms above the percolation point teams overlap substantially. The percolation point is the point at which the majority of units begin to touch. Below it, organizational units generally do not overlap, and above it organizational units generally do. Our formulation is general and allows for a wide range of interpretations that are consistent with the probabilistic formulation such as variations of goal-directed search, random realization of the economic benefit, biased distribution of information and teams in the space, information that is passed around the firm to find an expert, amongst others.}\label{gr:info space}
\end{figure}

We model how such a transition may occur with a simplified picture of how firms process information and extract benefits as in Figure~\ref{gr:info space}. We picture each record to be a point in a high-dimensional information space as indicated by the arrows. Each employee has some expertise, corresponding to a volume in the same space; this can be represented as a ``ball'' with a characteristic radius $r$. Then, we assume that firms extract some economic benefit $B$ from each piece of information if there is overlap between the expertise of an employee and the information content of a record. The chance that the record intersects with the expertise of the employee is the ratio of the size of the ball to the volume of the space, or the fraction $p$. Considering a firm with $N$ employees in a large information space, the probability that the piece intersects with the expertise of at least one employee is $1-(1-p)^{N}$. The result is exponential convergence to full coverage with employee number, or the prediction of a sudden turning point as a firm goes from a small size with a sparse covering to a large one with a dense covering of the information space.

The level at which useful information is extracted, however, is typically not relegated to a single employee but an organizational unit (e.g.~teams, branches, etc.), which may serve as a collection of expertise relatively independent of other units. Then, it is more accurate to consider a scaling relation of the organizational unit with the number of employees $N^b$ for a positive and sublinear exponent $b$, which implies that the probability of finding at least one unit with the right expertise is $1-(1-p)^{N^b}$, where $b=1$ as considered above corresponds to a firm with no organizational grouping. Simultaneously, firms are limited in the total amount of incoming information they can process regardless of overlap with expertise, a quantity that we model as a scaling with the number of employees of $N^a$. As before, the typical number of successful intersections between an information piece and an organizational unit yields a total economic benefit $B$. Assuming that cost of new information is proportional to its frequency (i.e.~it pays for itself),\footnote{If it were to pay more than for itself, we would anticipate that more information would be consumed and if it paid for less than the amount would be reduced. In other words, the marginal benefit should equal the marginal cost.} we would expect that $B$ is proportional to the total new information
\begin{align}
	I_{\rm tot} \sim B \sim N^a \left[1-(1-p)^{N^b}\right]. \label{eq:info benefit}
\end{align}

Eq~\ref{eq:info benefit} predicts three regimes. First, there is a regime in which each additional organizational unit contributes more or less independently to the expertise of the firm, maximizing the benefit of each organizational addition. Then, there is a relatively sharp, exponential turning point at which the organizational units suddenly saturate the space of information and when every new piece of information almost always intersects with some unit's expertise. Thereafter, benefits from filling the space are marginal, but the rate of new information is only determined by a universal rate $N^a$ at which organizations read. Since this is also the regime in which organizational units overlap in expertise, this is compatible with the observation that for large firms the dominant limit is not cumulative expertise but the fact that units must coordinate. Thus, Eq~\ref{eq:info benefit} presents a testable prediction for how the way that firms fill the space of information and the onset of a coordination problem manifest in the cost of reading. 

To fit the model to the data, we minimize the least-squares error on a logarithmic scale. While we have a sufficiently wide range of data to simply fit articles and sources, we are limited in the number of topics, which leads the curve to flatline at about $R\sim10^3$. A natural solution seems to be to fit to the part of the curve that comes before the flatline, but this leads to the problem of choosing the maximum values $R_{\rm max}$ below which to fit. Since there is no definitive point to which to restrict ourselves, we instead vary the cutoff $R_{\rm cutoff}$ a wide range. Helpfully, we find the errors to be large when we restrict ourselves to fit a range $10^2 \lesssim R \lesssim 10^3$ records, but they suddenly drop at larger $R$. There, we find a range in which the exponent $a$ that determines the extrapolated region stays within a narrow range. Finally, we find beyond $10^3 \lesssim R \lesssim 10^4$ that the exponent for the mean curve $a$ exceeds that for the maximum curve, which is physically impossible and a indication that the curves overfit the flatline (see Appendix~\ref{si sec:fitting} for more details). Thus, we find a natural fitting regime in which the extrapolation remains consistent while not overfitting the data cutoff.

Remarkably, our model matches closely the curves in Figure~\ref{gr:heaps}. Using the scaling relation between records $R$ and employees $N$ to replace $N$ with $R$, we fit the prediction in Eq~\ref{eq:info benefit} to the Heaps' plots in Figure~\ref{gr:heaps}. The data agree extraordinarily well with the mean and maximum curves predicted from the emergence of a coordination problem shown in the blue and orange curves, respectively. Importantly, our estimate for the exponent $b$, how team size scales with the nmber of employees, are all sublinear. For the mean curves, we find for articles, sources, and topics, $b\approx 3/10$, $b\approx1/3$, and $b\approx1/2$, respectively. For maximally diverse reading, we find $b\approx 3/10$, $b\approx2/5$, and $b\approx2/3$. Sensibly, the increasing values indicate that bigger collections of employees map onto larger aggregates of information.

The observation that larger groups matter for larger aggregations of information also manifests in the place at which the maximum variety curves peel away from the one-to-one line. When the inflection point is defined as the number of records at which the quantity in question first reaches 90\% of linear growth, it occurs at $R=1{,}645$ records for articles, $R=41$ records for sources, and $R=18$ records for topics. Using our measured scaling relations, the points correspond to publicly listed assets and annual sales of typically \$900 million and \$100 million, $\$80$ million and \$6 million, and \$40 million and \$3 million, respectively. That the variation of the inflection point maps to firms of different sizes suggests that firms may pass through critical sizes at which the amount of new information of a certain granularity cannot be processed in the same way.

\section*{Information intensive vs.~extensive growth}

As the final comparison, we consider how firm size relates to topic variety. As a start, we might anticipate three qualitatively different scenarios relating size such as assets $A$ with topic variety $T$, or the relation $A \sim T^\gamma$ for positive exponent $\gamma$. If firms were to focus on core competencies, we might expect superlinear scaling of $\gamma>1$ since that means firms tend to reinvest topics that they already read. Then, we would expect that the ratio of assets per topic $A/T \sim T^{\gamma-1}$ increases. In the linear case $\gamma=1$, we would find that a fixed unit of asset growth corresponds to the addition of a topic, or a siloed portrait of a firms such as might be expected from the separate acquisitions of a conglomerate that remain unintegrated. The final, sublinear case $\gamma<1$ would be unusual because it would suggest that assets are being divested from existing interests while more go into an enlargened set. Since larger firms tend to be older \cite{zhangScalingLaws2021}, the ratio of $A/T$ may reflect as aspect of strategy that we distinguish as either ``intensive'' (superlinear) or ``extensive'' (sublinear).

To determine $\gamma$, we must rely on the previous relations for assets $A$ as a function of records $R$ from Figure~\ref{gr:capital}, which was approximately $A\sim R^{2/3}$, and the predicted relation between topics $T$ and $R$ given by exponent $a$ measured in Figure~\ref{gr:heaps}. For firms of mean reading variety, we found $a\approx1/4$ and for firms of maximal variety we found $a\approx1/2$. Putting the relations together, we estimate for large firms the relations $A\sim T^{8/3}$ and $A\sim T^{4/3}$, respectively. Confidence intervals of 95\% on the values of $a$ range from $[0.05,0.34]$ and $[0.28,0.78]$ correspond to $\gamma$ ranging from $[0.07,0.50]$ and $[0.41,1.15]$; thus, they remain mostly confined to the superlinear regime, if not entirely. Overall, this is consistent with the picture of a firm that retains a fixed set of interests into which it reinvests assets, although they also allow for rare instances of alternative ``silo'' and extensive strategies. Taken together with our other findings, this observation indicates how the information dimension may reflect elements of organizational structure.

\begin{figure}\centering
	\includegraphics[width=.9\linewidth]{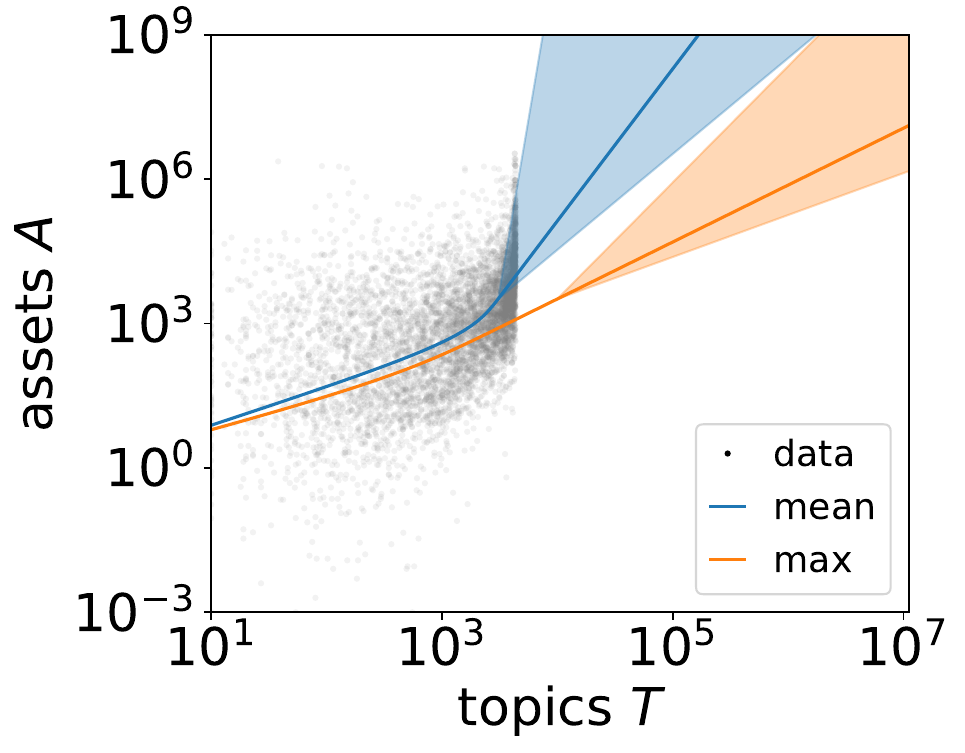}
	\caption{Predicted asset growth with topics from combining the information processing model from Figure~\ref{gr:heaps} and the scaling between assets and records (Table~\ref{si tab:exponents}). Gray points indicate public firms. A superlinear curve in the tail indicates increasing assets per topic read, whereas a sublinear curve decreasing assets per topic read, or an intensive vs.~extensive strategy, respectively. Linear scaling indicates proportional growth in assets with topic variety such as if investments were split equally across them. Best estimates for scaling in the tail are approximately $A\sim T^{8/3}$ for the mean and $A\sim T^{4/3}$ for the max, which both correspond to intensive strategies. Shaded regions indicate 95\% confidence intervals as defined in Appendix \ref{si sec:fitting}. Bottomost error bars extend to sublinear scaling, or roughly as $A \sim T^{8/9}$.}\label{gr:info strategies}
\end{figure}

\section*{Discussion}
How groups of biological organisms exchange information to coordinate individual components has been major area of interest in the study of collective behavior \cite{couzinSelfOrganizationCollective2003,rosenthalRevealingHidden2015,brushFamilyAlgorithms2013}. Yet, it is difficult to query the cognitive state of the individual, and a sophisticated experimental apparatus is crucial to control and track the information that individuals are receiving and generating \cite{Noy:2011kr,rosenthalRevealingHidden2015,Minderer:2016je,Stowers:2017ia,leeAudioCues2019}. In organizations, much transmitted information is recorded and its content understandable; instead, the scale and complexity of the information is little understood quantitatively. Thus, the firm presents a complementary opportunity to study information flows. We leverage such information at a wide scale to study on the amount of information that organizations, primarily firms, consume and focus on how such information is connected to firm size.

We first show a sublinear relationship between the size and the information footprint of a firm in Figures~\ref{gr:capital} and \ref{gr:power laws}. The main implication of the sublinear scaling is that the effective cost of an additional unit of information shrinks, signaling an ``economy of scale.'' This observation aligns with the intuition that the tools of the knowledge economy such as information technology use enhance firm productivity \cite{stirohInformationTechnology2002,blackWhatDriving2004}. As part of this, the typical large firm accesses more {\it records} and consumes more {\it articles} per employee, which implies that its employees on the whole are more productive at processing elementary information. The sublinear scaling leads to an information consumption inequality that is an exaggerated version of the economic inequality between firms given by the classic finding of Zipf's law in US Census data \cite{axtellZipfDistribution2001}. We validate this by measuring the exponents of power-law tails in the distributions of firm information consumption $\alpha$, finding an exaggerated tail for records and articles of $\alpha<2$. Sources, in contrast, show nearly linear scaling, and we confirm that the power law tail is Zipfian, or $\alpha=2$. Flipping the axes around by taking information as the driver, the inverse perspective is of a superlinear increase in information required to ``power'' unit economic growth, or one of diminishing returns. The two perspectives, while compatible with the same scaling laws, correspond to different causal mechanisms for the drivers of productivity in firms.

The change in the scaling from sublinear to linear as we change information granularity is curious. This is summarized in the scaling exponents $\beta$ in Figure~\ref{gr:power laws}, which indicate the strongest economy of scale for records and articles and the weakest for topics. 
One possible reason for the variation is that processing diverse information at larger scales is more complicated. Gaining something out of a whole new topic may require restructuring of the firm such as adding a division or changing the corporate mission, whereas reading a new article on the same topic is trivial. This would mean that firm growth in information space at the coarsest levels is more tied to economic growth, echoing the role of intangible aspects in determining firm costs \cite{blackHowCompete2001}. The observations suggest two different ways in which information costs may reflect firm growth, either through an increasing economy of scale or reflecting the demands of diversification.

As a second observation, we show that the volume of firm online reading is limited in its variety because the number of read articles, sources, and topics scales sublinearly with volume in the Heaps' plots of Figure~\ref{gr:heaps}. Even the most diversely read firms are redundant readers above a critical size, and the critical size depends on information granularity. By considering how the expertise of organizational units in a firm tile the space of information (Figure~\ref{gr:info space}), we predict the emergence of a transition when the organizational units in a firm begin to overlap in expertise. Overlap means the point at which coordination or conflict between units becomes an issue. This model yields a prediction that fits the data remarkably well, indicating that organizational constraints may leave traces in the information footprint. 

Finally, we connect reading variety to firm size by using the scaling relations found in the previous analyses Figure~\ref{gr:info strategies}. Inspired by the qualitative differences that could emerge for the scaling relations relating the two properties, we predict three types of firm strategy that would correspond to each one. The differences are summarized in the ratio of assets to topics $A/T$ that either grows, stays constant, or shrinks with firm size. We find that nearly the entire range compatible with the fit to data corresponds to an increasing ratio, or an intensive strategy, consistent with other measures \cite{hobergScopeScale2023}. This is in contrast with the economy of scale for records and articles. The difference may reflect the fact that expanding the interests of a firm is not as simple as reading another article, but it may involve a structural cost (e.g.~hiring the right employees and setting up management structures) reflected in processing higher levels of information. 

In each of these macroscopic trends, we find indications that organizational units, not the individual employees, play a role in information consumption. These echo the classic ideas that organizational structures are crucial for the ``absorptive capacity'' of a firm \cite{cohenAbsorptiveCapacity1990}, the way knowledge is stored or exploited \cite{kogutKnowledgeFirm1992}, and the role of teams \cite{wuLargeTeams2019} and organizational structure \cite{stanleyScalingBehaviour1996} on performance. Here, we provide another piece of the puzzle by measuring aspects of information consumption that elucidate the information behavior of employees. Information use by individuals follows a parallel line of work in biological collectives, where we have begun to connect individual-level information exposure and cognition to group-level capabilities \cite{dedeoInductiveGame2010a,kaoCollectiveLearning2014,leeCollectiveMemory2017,shishkovSocialInsects2022a}. How information consumption captures the capacity or interaction-structure in firms remains an open question. As a start, how firms vary in such organization as would be reflected in information use suggests one way to distinguish systems from one another or to highlight unusual ones. Large deviations from patterns extracted over many millions of firms are likely to represent surprising activity that demands further attention. In this sense, our work suggests a way forward for understanding why and how firms use information and the principles that organize information processing in social and biological systems.

\section*{Acknowledgements}
We thank the Complexity Science Hub Theory Group including Tuan Pham, Jan Korbel, and Stefan Thurner and others including Ernesto Ortega, Chris Kempes, and Geoffrey West for useful discussions. We thank Maria del Rio-Chanona for comments on a previous draft. E.D.L.~acknowledges funding from BMBWF, HRSM 2016 (Complexity Science Hub Vienna) and the Austrian Science Fund under grant number ESP 127-N. F.N.~acknowledges financial support from the Austrian Research Agency (FFG), project \#873927 (ESSENCSE).

\section*{Data availability}
Anonymized code for analysis can be made available upon request, but data presented are proprietary, and we must consider requests for data access for reproducibility on a case-by-case basis.

\bibliography{refs,frefs}

\begin{thebibliography}{10}

\bibitem{couzinEffectiveLeadership2005}
Couzin ID, Krause J, Franks NR, Levin SA (2005) Effective leadership and
  decision-making in animal groups on the move.
\newblock {\em Nature} 433(7025):513--516.

\bibitem{balleriniInteractionRuling2008}
Ballerini M, et~al. (2008) Interaction ruling animal collective behavior
  depends on topological rather than metric distance: {{Evidence}} from a field
  study.
\newblock {\em Proc Natl Acad Sci USA} 105(4):1233--1237.

\bibitem{rosenthalRevealingHidden2015}
Rosenthal SB, Twomey CR, Hartnett AT, Wu HS, Couzin ID (2015) Revealing the
  hidden networks of interaction in mobile animal groups allows prediction of
  complex behavioral contagion.
\newblock {\em Proc. Natl. Acad. Sci. U.S.A.} 112(15):4690--4695.

\bibitem{hartnettHeterogeneousPreference2016}
Hartnett AT, Schertzer E, Levin SA, Couzin ID (2016) Heterogeneous
  {{Preference}} and {{Local Nonlinearity}} in {{Consensus Decision Making}}.
\newblock {\em Phys. Rev. Lett.} 116(3):038701.

\bibitem{brushFamilyAlgorithms2013}
Brush ER, Krakauer DC, Flack JC (2013) A {{Family}} of {{Algorithms}} for
  {{Computing Consensus}} about {{Node State}} from {{Network Data}}.
\newblock {\em PLoS Comput Biol} 9(7):e1003109.

\bibitem{wuchty2007increasing}
Wuchty S, Jones BF, Uzzi B (2007) The increasing dominance of teams in
  production of knowledge.
\newblock {\em Science} 316(5827):1036--1039.

\bibitem{nonakaKnowledgecreatingCompany1995}
Nonaka I, Takeuchi H (1995) {\em The {{Knowledge-creating Company}}: {{How}}
  the {{Japanese Companies Create}} the {{Dynamics}} of {{Innovation}}}.
\newblock ({Oxford University Press}).

\bibitem{bhattLanguageBasedMethod2022}
Bhatt AM, Goldberg A, Srivastava SB (2022) A {{Language-Based Method}} for
  {{Assessing Symbolic Boundary Maintenance}} between {{Social Groups}}.
\newblock {\em Sociol. Methods Res.} 51(4):1681--1720.

\bibitem{simonInformationProcessing1979}
Simon HA (1979) Information {{Processing Models}} of {{Cognition}}.
\newblock {\em Ann. Rev. Psychol.} 30:363--96.

\bibitem{katzInvestigatingNot1982}
Katz R, Allen TJ (1982) Investigating the {{Not Invented Here}} ({{NIH}})
  syndrome: {{A}} look at the performance, tenure, and communication patterns
  of 50 {{R}} \& {{D Project Groups}}.
\newblock {\em R \& D Management} 12(1):7--20.

\bibitem{katzInferringStructure2011}
Katz Y, Tunstr{\o}m K, Ioannou CC, Huepe C, Couzin ID (2011) Inferring the
  structure and dynamics of interactions in schooling fish.
\newblock {\em Proc. Natl. Acad. Sci. U.S.A.} 108(46):18720--18725.

\bibitem{simonNewDevelopments1962}
Simon HA (1962) New {{Developments}} in the {{Theory}} of the {{Firm}}.
\newblock {\em The American Economic Review} 52(2):1--15.

\bibitem{cohenAbsorptiveCapacity1990}
Cohen WM, Levinthal DA (1990) Absorptive {{Capacity}}: {{A New Perspective}} on
  {{Learning}} and {{Innovation}}.
\newblock {\em ASQ} 35(1):128--152.

\bibitem{granovetterStrengthWeak1973}
Granovetter MS (1973) The {{Strength}} of {{Weak Ties}}.
\newblock {\em AJS} 78(6):1360--1380.

\bibitem{burtStructuralHoles2004}
Burt RS (2004) Structural {{Holes}} and {{Good Ideas}}.
\newblock {\em Am. J. Sociol.} 110(2):349--399.

\bibitem{strogatzExploringComplex2001}
Strogatz SH (2001) Exploring complex networks.
\newblock {\em Nature} 410(6825):268--276.

\bibitem{albertStatisticalMechanics2002}
Albert R, Barab{\'a}si AL (2002) Statistical mechanics of complex networks.
\newblock {\em Rev. Mod. Phys.} 74(1):47--97.

\bibitem{wattsCollectiveDynamics1998}
Watts DJ, Strogatz SH (1998) Collective dynamics of `small-world' networks.
\newblock {\em Nature} 393:440--442.

\bibitem{deming2017growing}
Deming DJ (2017) The growing importance of social skills in the labor market.
\newblock {\em The Quarterly Journal of Economics} 132(4):1593--1640.

\bibitem{neffke2019value}
Neffke FM (2019) The value of complementary co-workers.
\newblock {\em Science advances} 5(12):eaax3370.

\bibitem{girvanCommunityStructure2002}
Girvan M, Newman MEJ (2002) Community structure in social and biological
  networks.
\newblock {\em Proc. Natl. Acad. Sci. U.S.A.} 99(12):7821--7826.

\bibitem{hallRecentResearch2012}
Hall BH, Harhoff D (2012) Recent {{Research}} on the {{Economics}} of
  {{Patents}}.
\newblock {\em Annu. Rev. Econ.} 4(1):541--565.

\bibitem{jaffeMeaningPatent2000}
Jaffe A, Trajtenberg M, Fogarty M (2000) The {{Meaning}} of {{Patent
  Citations}}: {{Report}} on the {{NBER}}/{{Case-Western Reserve Survey}} of
  {{Patentees}}, ({National Bureau of Economic Research}, {Cambridge, MA}),
  Technical Report w7631.

\bibitem{younInventionCombinatorial2015}
Youn H, Strumsky D, Bettencourt LMA, Lobo J (2015) Invention as a combinatorial
  process: Evidence from {{US}} patents.
\newblock {\em J. R. Soc. Interface} 12(106):20150272.

\bibitem{kwanDoesInternet2019}
Kwan A, Zhu C (2019) Does {{Internet Research Activity}} by {{Sophisticated
  Investors Lead}} to {{Heightened Adverse Selection}}?
\newblock {\em SSRN Journal}.

\bibitem{kwanInstitutionalInvestor2022}
Kwan A, Liu Y, Matthies B (2022) Institutional {{Investor Attention}}.

\bibitem{hobergConglomerateIndustry2018}
Hoberg G, Phillips G (2018) Conglomerate {{Industry Choice}} and {{Product
  Language}}.
\newblock {\em Management Science} 64(8):3735--3755.

\bibitem{hobergTextBasedNetwork2016}
Hoberg G, Phillips G (2016) Text-{{Based Network Industries}} and {{Endogenous
  Product Differentiation}}.
\newblock {\em Journal of Political Economy} 124(5).

\bibitem{axtellZipfDistribution2001}
Axtell RL (2001) Zipf {{Distribution}} of {{U}}.{{S}}. {{Firm Sizes}}.
\newblock {\em Science} 293(5536):1818--1820.

\bibitem{teeceUnderstandingCorporate1994}
Teece DJ, Rumelt R, Dosi G, Winter S (1994) Understanding corporate coherence.
\newblock {\em Journal of Economic Behavior \& Organization} 23(1):1--30.

\bibitem{hidalgoEconomicComplexity2021}
Hidalgo CA (2021) Economic complexity theory and applications.
\newblock {\em Nat Rev Phys} 3(2):92--113.

\bibitem{zhangScalingLaws2021}
Zhang J, Kempes CP, Hamilton MJ, West GB (2021) Scaling laws and a general
  theory for the growth of companies.
\newblock {\em arXiv:2109.10379 [physics]}.

\bibitem{clausetPowerLawDistributions2009}
Clauset A, Shalizi CR, Newman MEJ (2009) Power-{{Law Distributions}} in
  {{Empirical Data}}.
\newblock {\em SIAM Rev.} 51(4):661--703.

\bibitem{heapsInformationRetrieval1978}
Heaps HS (1978) {\em Information Retrieval, Computational and Theoretical
  Aspects}.
\newblock ({Academic Press}).

\bibitem{couzinSelfOrganizationCollective2003}
Couzin ID, Krause J (2003) Self-{{Organization}} and {{Collective Behavior}} in
  {{Vertebrates}} in {\em Advances in the {{Study}} of {{Behavior}}}.
\newblock ({Elsevier}) Vol.{}~32, pp. 1--75.

\bibitem{Noy:2011kr}
Noy L, Dekel E, Alon U (2011) The mirror game as a paradigm for studying the
  dynamics of two people improvising motion together.
\newblock {\em PNAS} 108(52):20947--20952.

\bibitem{Minderer:2016je}
Minderer M, Harvey CD, Donato F, Moser EI (2016) Neuroscience: {{Virtual}}
  reality explored.
\newblock {\em Nature} 533(7603):324--325.

\bibitem{Stowers:2017ia}
Stowers JR, et~al. (2017) Virtual reality for freely moving animals.
\newblock {\em Nat Meth} 14(10):995--1002.

\bibitem{leeAudioCues2019}
Lee ED, Esposito E, Cohen I (2019) Audio cues enhance mirroring of arm motion
  when visual cues are scarce.
\newblock {\em J. R. Soc. Interface} 16(154):20180903.

\bibitem{stirohInformationTechnology2002}
Stiroh KJ (2002) Information {{Technology}} and the {{U}}.{{S}}. {{Productivity
  Revival}}: {{What Do}} the {{Industry Data Say}}?
\newblock {\em The American Economic Review} 92(5):1559--1576.

\bibitem{blackWhatDriving2004}
Black SE, Lynch LM (2004) What's {{Driving}} the {{New Economy}}? {{The
  Benefits}} of {{Workplace Innovation}}.
\newblock {\em The Economic Journal} p.~20.

\bibitem{blackHowCompete2001}
Black SE, Lynch LM (2001) How to {{Compete}}: {{The Impact}} of {{Workplace
  Practices}} and {{Information Technology}} on {{Productivity}}.
\newblock {\em The Review of Economics and Statistics} 83(3):434--445.

\bibitem{hobergScopeScale2023}
Hoberg G, Phillips GM (2023) Scope, {{Scale}} and {{Concentration}}: {{The}}
  21st {{Century Firm}}.

\bibitem{kogutKnowledgeFirm1992}
Kogut B, Zander U (1992) Knowledge of the {{Firm}}, {{Combinative
  Capabilities}}, and the {{Replication}} of {{Technology}}.
\newblock {\em Organ. Sci.} 3(3):383--397.

\bibitem{wuLargeTeams2019}
Wu L, Wang D, Evans JA (2019) Large teams develop and small teams disrupt
  science and technology.
\newblock {\em Nature} 566(7744):378--382.

\bibitem{stanleyScalingBehaviour1996}
Stanley MHR, et~al. (1996) Scaling behaviour in the growth of companies.
\newblock {\em Nature} 379(6568):804--806.

\bibitem{dedeoInductiveGame2010a}
DeDeo S, Krakauer DC, Flack JC (2010) Inductive {{Game Theory}} and the
  {{Dynamics}} of {{Animal Conflict}}.
\newblock {\em PLoS Comput Biol} 6(5):e1000782.

\bibitem{kaoCollectiveLearning2014}
Kao AB, Miller N, Torney C, Hartnett A, Couzin ID (2014) Collective
  {{Learning}} and {{Optimal Consensus Decisions}} in {{Social Animal Groups}}.
\newblock {\em PLoS Comput Biol} 10(8):e1003762.

\bibitem{leeCollectiveMemory2017}
Lee ED, Daniels BC, Krakauer DC, Flack JC (2017) Collective memory in primate
  conflict implied by temporal scaling collapse.
\newblock {\em J. R. Soc. Interface} 14(134):20170223.

\bibitem{shishkovSocialInsects2022a}
Shishkov O, Peleg O (2022) Social insects and beyond: {{The}} physics of soft,
  dense invertebrate aggregations.
\newblock {\em Collective Intelligence} 1(2):263391372211237.

\end{thebibliography}

\end{document}


\maketitle

\renewcommand{\thefigure}{S\arabic{figure}}
\renewcommand{\thetable}{S\arabic{table}}

\appendix
\section{Data preprocessing}\label{si sec:data}
The data set consists of access records of news content that is within the universe of publishers monitored by a firm specializing in ``intent'' data, hereafter called the ``Data Partner.'' As we specify in reference \cite{kwanDoesInternet2019}, 

\begin{quotation}
`Intent' refers to a recent strain of data analytics aiming to gauge a prospective business customer’s buying interest based on patterns of reading on the internet. If a given business customer’s reading on a particular topic increases at a rate relative to a baseline, one might presume the business customer has a greater `intention' of transacting in a related service as this increase in reading is indicative of the research that occurs prior to a purchase.
\end{quotation}

\begin{figure}[!b]\centering
	\includegraphics[width=\linewidth]{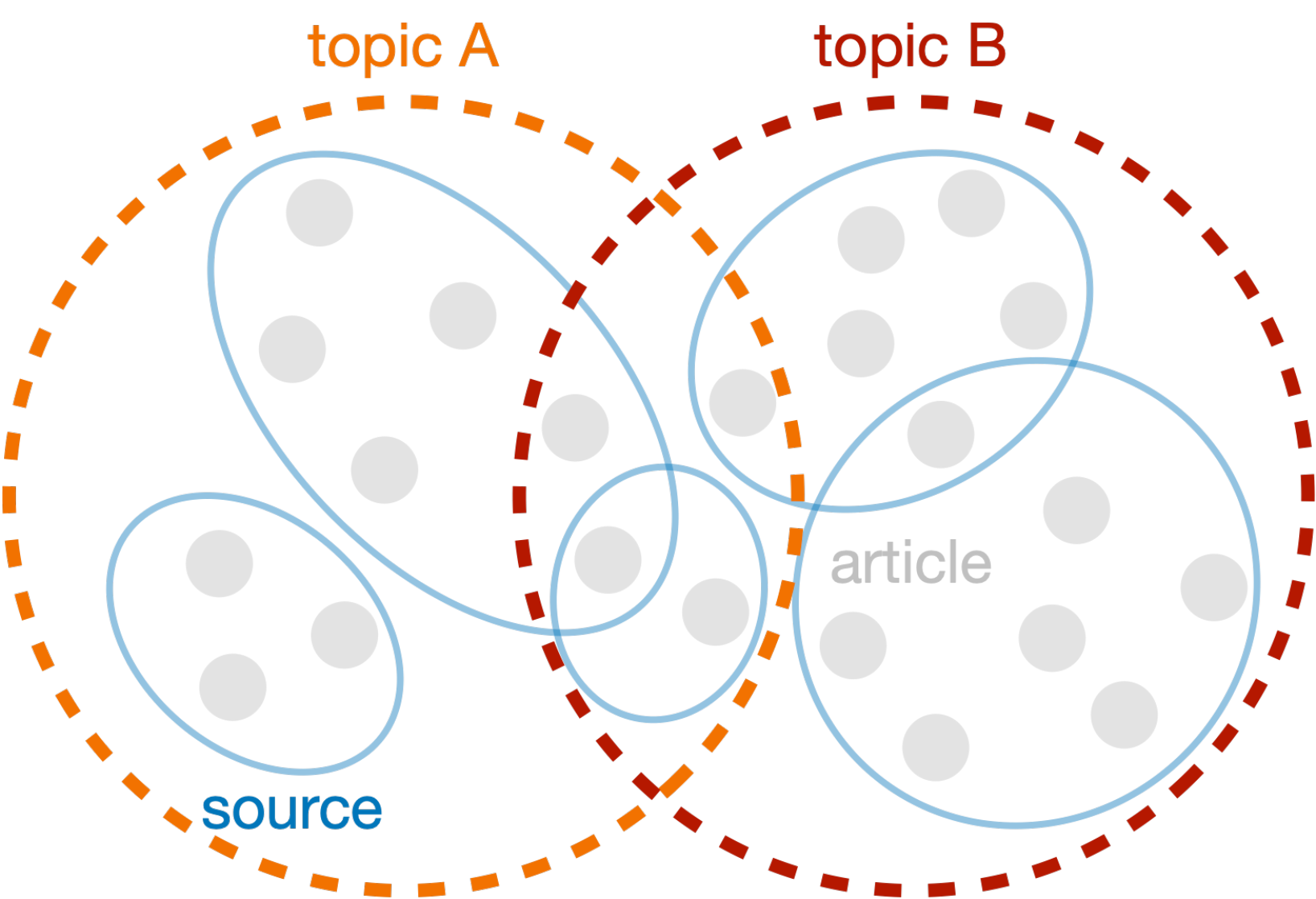}
	\caption{Information falls across multiple layers of resolution from articles (gray circles) provided by on different pages at a publisher's website, or sources (blue circles), which can be related by topic (dashed circles). Articles form the most unstructured view of the data and each article can belong to multiple sources or topics that form different scales for analysis. Here, we consider a copy of the same article on a different content provider as a separate article.}\label{gr:scales}
\end{figure}

Each record, in principle, does not identify who is accessing the content, only the origin of the request, or the IP address, as well as metadata associated with the visitor, known as a ``cookie.'' Cookies allow a user to be observed through different visitor sessions. Using methods standard in the digital marketing industry, the Data Partner links various website visits to a given firm. The Data Partner analyzes the textual content and uses proprietary topic model implementation to extract a set of hand-labeled, customer-relevant topics associated with each page of content. Thus, we have a database of which devices are accessing what and information about the content in question. 

For the topic model, the Data Partner runs a binary classification model which seeks to filter out content which is not relevant to business, trained on a set of articles from sports, adult, and entertainment websites versus a set of articles from well-known business publishers. After filtering the content which is likely to discuss business-relevant content, the Data Partner runs a multi-label model which calculates the article's topic relevancy to the current taxonomy of topics, which during our sample period was around 4{,}000 topics (not all are commercially sold). The intersection of content which can be mapped to a firm, and content which can be mapped to a business, is about 10-15\% of the total content observed by the Data Partner (consisting of over one billion article read events per day). Thus, although it is not clear that all information consumed by the organization is done so for productive purposes, the filters applied toward the content and visitors suggests that information consumption events observed in this study contain the subset of observations most likely to come from work-related visitors and work-related content.

To provide some context, here we provide a breakdown of articles by publisher type in terms of their IAB category (a common taxonomy for content developed by the Interactive Advertising Bureau, \url{https://iabtechlab.com/standards/content-taxonomy/}) as well as the AlexaRank. While imperfect, these are two popular classifications applicable to all websites across the Internet. They suggest that the plurality of the information observed in this study come from major news and business publishers but the content altogether is diverse.

\begin{table}[t]\centering
\caption{IAB classifications of the top-level domains (provided by Data Partner) breaking down the percentages of articles across the publisher types and publisher web traffic rank, per AlexaRank, from the year 2019.}\label{si tab:publisher_alexarank}

\begin{tabular}{@{\extracolsep{5pt}} ccc} 

\hline \\[-1.8ex] 
IAB & \% Total & IAB name \\ 
\hline \\[-1.8ex] 
IAB12 & $21.77$ & News \\ 
IAB3 & $8.81$ & Business \\ 
IAB13 & $8.41$ & Personal Finance \\ 
IAB19 & $7.79$ & Technology, Computing \\ 
IAB9 & $6.37$ & Hobbies, Interests \\ 
IAB25 & $5.48$ & Non-Standard Content \\ 
IAB1 & $5.27$ & Arts, Entertainment \\ 
IAB5 & $4.22$ & Education \\ 
IAB22 & $4.07$ & Shopping \\ 
IAB17 & $3.68$ & Sports \\ 
IAB15 & $3.39$ & Science \\ 
IAB7 & $3.08$ & Health/Fitness \\ 
IAB8 & $2.45$ & Food / Drink \\ 
IAB11 & $2.37$ & Law, Gov’t, Politics \\ 
IAB20 & $1.88$ & Travel \\ 

\hline \\[-1.8ex] 
\end{tabular}

\begin{tabular}{@{\extracolsep{5pt}} cc} 
\\[-1.8ex]\hline 

AlexaRank & \% Total Reading \\ 
\hline \\[-1.8ex] 
$[1-1000]$ & $27.87$ \\ 
$[1000-10000]$ & $17.59$ \\ 
$[25000-50000]$ & $7.32$ \\ 
$[50000-500000]$ & $5.56$ \\ 
$[500000-10^6]$ & $15.91$ \\ 
$ > 10^6$ & $25.75$ \\ 

\hline \\[-1.8ex] 
\end{tabular} 
\end{table}

\begin{figure}[t]\centering
	\includegraphics[width=.8\linewidth]{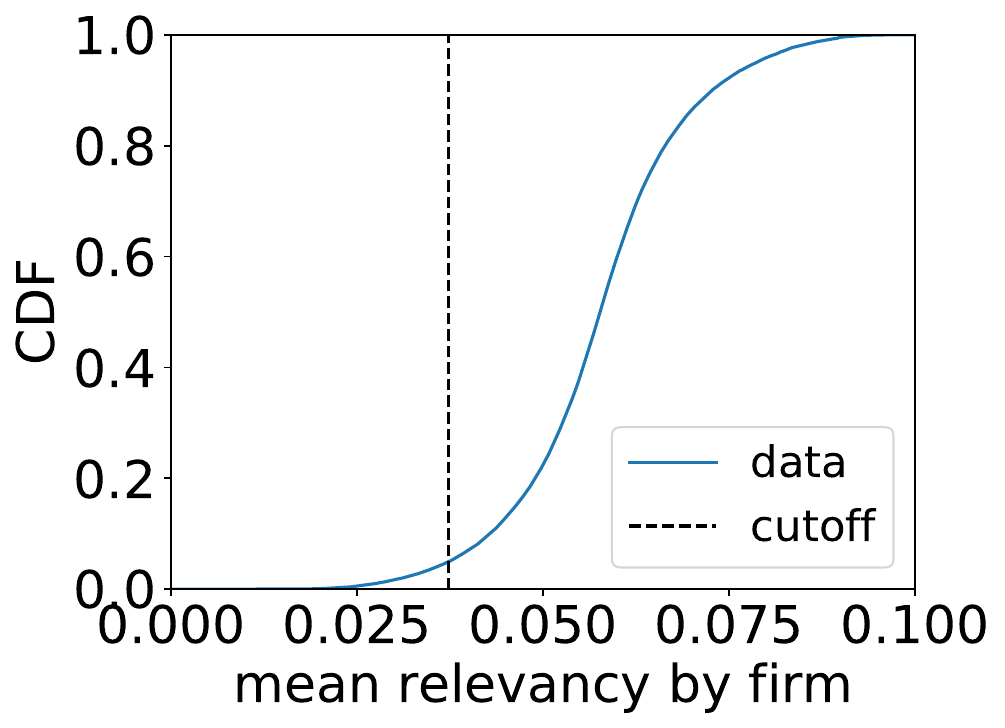}
	\caption{Cumulative distribution function (CDF) over firms of the firm-averaged relevancy scores. We use a 95th percentile threshold to remove firms whose reading habits are not well captured by the topic modeling (dashed black line).}\label{si gr:relevancy cutoff}
\end{figure}

The data is proprietary and the analyses are run on big data sets, so we cannot feasibly rerun our own preprocessing. However, we can take some steps to mitigate potential limitations of the data set. 
\begin{enumerate}
	\item One consideration is that the topic list is updated over time to account for changing data sources and customer interests. To limit this time-varying effect, we focus on a two-week period between June 10, 2018 and June 23, 2018 in the time zone GMT, for which we do not have any reason to believe is unusual relative to other points in the data set, which represents a large number of labeled topics, and that does not have changes in topic preprocessing. Within this time frame, the preprocessed data consists of 4{,}338 unique topics with more than 3.5 million firms as identified by their domains, and day-to-day statistics are similar.
	\item Some of the extracted topics may not describe well the content at hand, so we impose a lower threshold on the topic relevance value. We first calculate the typical relevancy over the set of all topics for each firm for a given day. When firm average relevancy falls in the bottom 5th percentile, we remove it from consideration as we show in Figure~\ref{si gr:relevancy cutoff}. This allows us, in our cross-firm comparison, to only consider firms that are well-represented within our given, bounded universe of topics.
	\item We exclude Amazon because it is a clear outlier by over an order of magnitude in terms of records, which most likely indicates that customer behavior was incorrectly flagged as employee behavior such as through Amazon Web Services. This is also a possible source of error in other cloud or internet service provider firms like Comcast and Microsoft, but there is no clear indication that they are outliers as is the case with Amazon.
\end{enumerate}

\begin{figure}\centering
	\includegraphics[width=\linewidth]{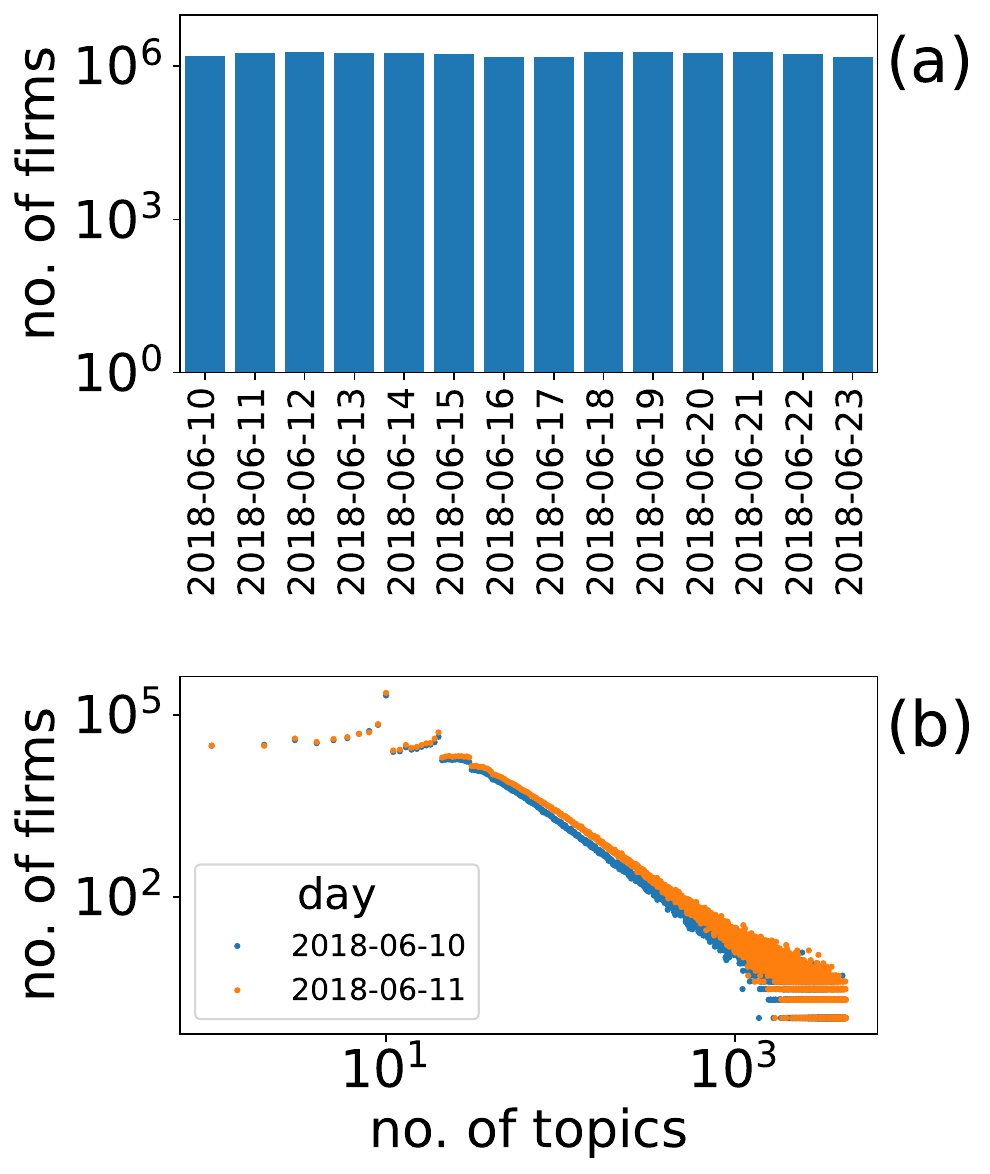}
	\caption{(a) Number of firms in database per day beforing filtering out firms with low mean topic-relevancy and subsampling. (b) Distribution of firms by number of topics across all queries on June 10, 2018 appears to be a scaled version of June 11.}\label{si gr:topic stats}
\end{figure}

Once the above steps have been taken, we have a selection of firms with some basic statistics plotted in Figure~\ref{si gr:topic stats}. Thus, we are considering roughly $10^6$ firms per day and the distribution of unique topics accessed by any given firm displays a power law tail.

Besides businesses, the data set contains government agencies, non-governmental organizations, academic institutions, and entities registered with a domain. While we could filter out some of the organizations, it is unclear if such pruning can be done in a consistent way across all countries especially when considering that firms in many countries are closely tied to the government such as Saudi Arabia's Aramco and Russia's Gazprom, to name a couple. As a check of how large an effect government organizations might have on the data set, we took an extensive list of 9{,}225 US government and government-affiliated URLs from \url{https://github.com/GSA/govt-urls} in addition to top-level ``.gov'' and ``.mil'' domains to find that of over 2.8 million domains, only 8{,}477 were US government affiliated. In terms of records, they constitute less than 4\% of the data. As for academic institutions, we consider them to be firms for the purposes of our analysis, but domains with a .edu suffix (either alone or followed by country-specific domain) also constitute about 9{,}152 and about 13\% of the records. Thus, the vast majority of the data we analyze corresponds to non-government, non-educational entities, and there is little indication that government agencies are unusual in our analyses.

\section{Fitting size scaling with reading volume}
In Figure~\ref{gr:capital}, we compare various measures of firm size $Y$ with the amount that firms read in terms of records $R$ in the data set to establish a sublinear relationship. It is natural, however, to consider the converse relationship of how reading volume scales with measures of firm size. This presents two different ways of considering fitting the data to obtain a scaling relationship, where the ``independent'' variable is either firm size or reading volume. It is not in general expected that the fits resolve the same scaling relationship. In other words, it is not necessarily the case that our finding $Y=AR^b$ when fit with the reversed axes returns $R=(Y/A)^{1/b}$ with the same fit parameters $A$ and $b$ as would be the case under an invertible scaling relation.

One solution is to use a symmetric cost function that accounts for logarithmic errors along both the $x$ and $y$ axes simultaneously, guaranteeing an invertible relationship. A more sophisticated technique would be total factor correlation, which accounts for the principle dimension through which the data passes \cite{}. We do not rely on either of the two techniques because there is clear asymmetry in the fit error distributions that is attributable to known biases in the reading data and we do not have a good model of the corresponding error distribution.

\begin{figure}\centering
    \includegraphics[width=\linewidth]{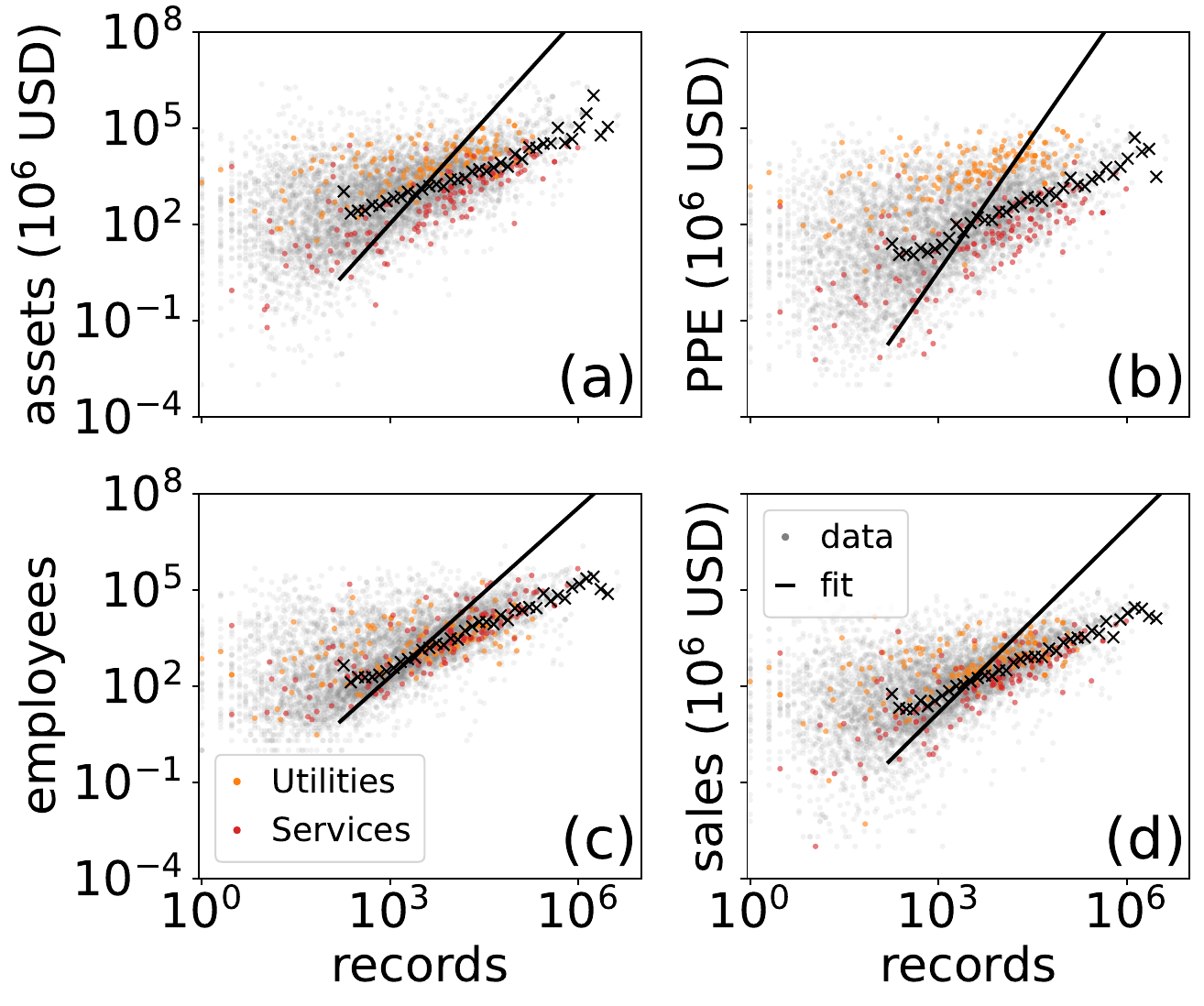}
    \caption{Relationship between firm size measures and reading volume $R$. Logarithmic regressions using least-squares but with errors along the $x$-axis are shown as black lines. Medians along the measures of firm size are shown as black x's.}\label{si gr:reversed cap fit}
\end{figure}

\begin{figure}\centering
	\includegraphics[width=.8\linewidth]{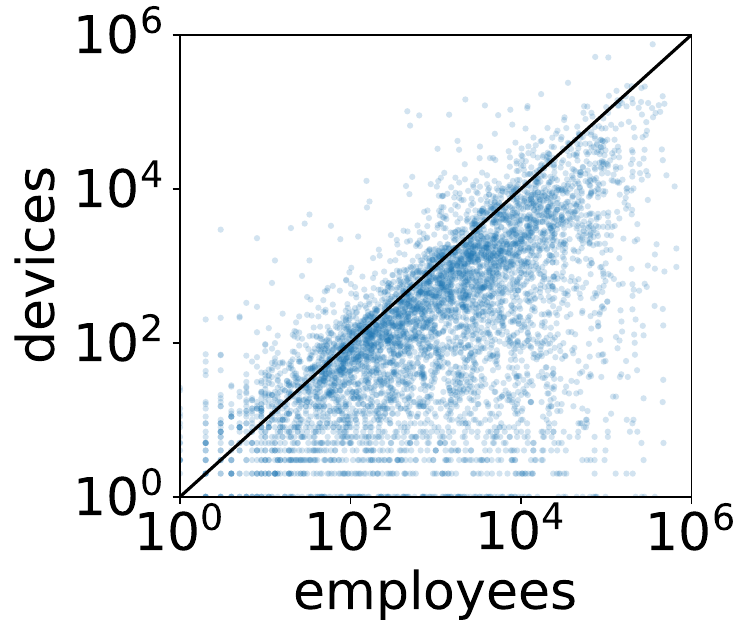}
	\caption{Number of devices per employee estimates from COMPUSTAT for the quarter ending on 2018-07-01. Black line indicates 1:1. There is a cloud of devices that follow a linear scaling, indicating that the number of devices, for a majority of firms, is proportional to the number of employees.}\label{si gr:devices per emp}
\end{figure}

As an example of what happens when we use firm measures as the independent variable is shown in Figure~\ref{si gr:reversed cap fit}. Whereas the regression that we show in the main text --- using logarithmic least-squares along the $y$-axis --- cuts right through the median data, a regression along the other axis does not. This reflects the fact that the reading coverage is uneven over firms, so there are many firms large and small for which coverage is sparse (see Figure~\ref{si gr:devices per emp}). This means that once we condition on firm size, and especially for large firms where we have few data points, skewed observations matter inordinately for the mean (note that conditioning on firm size is separate from averaging over firms with a large number of records as we do in our later analysis). In contrast, firm size measures from COMPUSTAT are complete and public firms must abide by the standardized accounting and disclosure regulations in place. Reflecting the comprehensive nature of firm sizes measures, we recover error distributions that are much closer to normal and pass through the median data points for firm size distributions, indicating that a logarithmic least-squares fit is tenable. Thus, we perform our fits to the relationship between firm size and reading only considering the errors for which we can be reasonably confident as being well-characterized with a simple fitting cost.

\section{Heaps model specification}\label{si sec:heaps}
As a hypothesis for how firms process the incoming flow of information, we picture a shared information space in which articles and organizational units in the firm live. The assumptions are that the intersection of an article (a point in the high-dimensional information space) with the radius of expertise of an organizational unit leads to a benefit for the firm. The prediction made by this model is summarized in the main text in Eq~\ref{eq:info benefit}. 

In order to fit our model to the data, we minimize the squared logarithmic distance to the data points that are shown in Figure~\ref{gr:heaps}. For articles and sources, the procedure is straightforward because we are unconcerned about saturating the full number of articles or sources in the data set under the period of study, so we fit to the variety count at each logarithmic bin in Figure~\ref{gr:heaps} as long as there are at least five observed firms and the firm with the minimum number of reads has more than one record (an indication of undersampling). For topics, however, fitting to the full range of records would mean that we are also fitting to the point where the finite number of observed topics caps the relationship between records and topics. 

To minimize the impact of the cutoff, we fit only to a limited regime far before the curve flattens out at large $R$ for topics. The limited regime also means, however, that the fit is much more sensitive to small aberrations in the data at the end of small firms, which are noisier and would be more strongly affected by bots. We solve this problem in two steps. First, we do not fit to the entire curve but consider a principled upper cutoff at the turning point that we identify earlier 90\% of the 1:1 one, or $R\leq 178$. As a result, however, the fit jumps to fractional values for small values of $R$, which is not physically possible. To solve this problem, we increment the lower cutoff until we recover a fit that goes through the data points and remains greater than unity. We find that for the mean curve (blue), we are limited to the range $11\leq R\leq 178$ and for the maxima (orange) $1\leq R \leq 178$. Note that the curvature of the data in the Heaps' plots are different from that of a null model where firms are randomly sampling from the set of articles, sources, and topics that we have in the data $R$ times. In the latter case, the mean reading variety curves run almost completely along the 1:1 line. The fact that firms are systematically re-reading items is reflected in the nontrivial exponents that we find for team-size scaling $b$ and the limits to large-firm reading $a$.
Thus, the fits to the Heaps' curves for articles and sources indicate that our simple model agrees closely with the data and extrapolates the scaling of topic variety to large firms, where the data are inadequate.

\section{Sector-by-sector fits}
\begin{figure*}\centering
    \includegraphics[width=.8\linewidth]{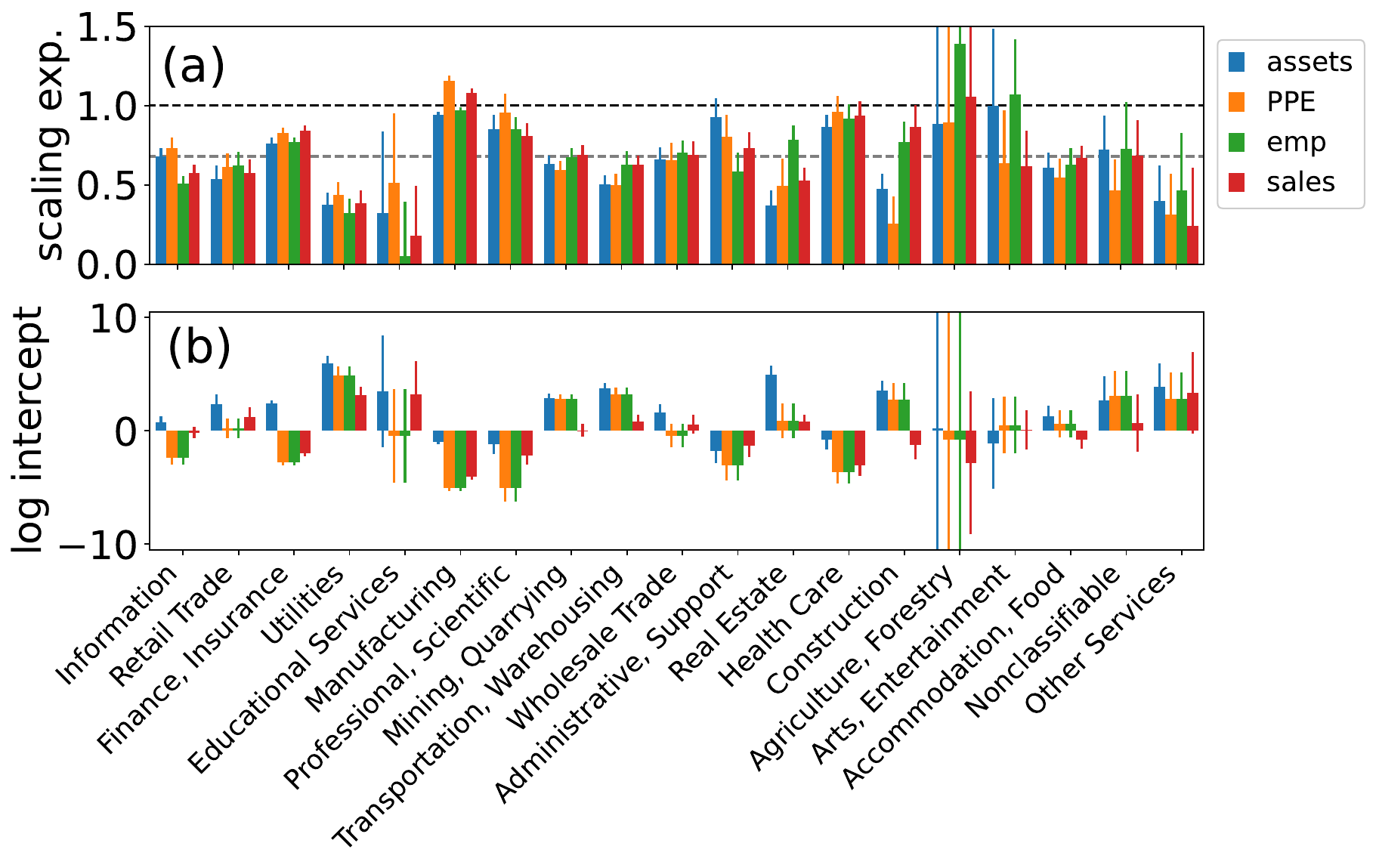}
    \caption{Scaling model (Eq~\ref{eq:scaling}) fits to each NAICS sector separately for (a) exponents and (b) intercepts. For each sector, we show the first two words of the sector name along the $x$-axis. In panel a, we show a black, dashed line at linearity, and a gray, dashed line at an exponent value of 0.68, the value measured when fitting over all firms at once.}\label{si gr:sector fits}
\end{figure*}

We fit the scaling relations between firm size measures and records for each NAICS sector that we have represented in the database. We have taken the primary NAICS classification code for each firm as shown in Figure~\ref{si gr:sector fits}. While there is variation across the sectors, we find that the scaling exponents in panel a are consistently sublinear. In the few cases where they are not, the bootstrapped error bars indicate that we cannot be sure of the precise value.

\section{Information space model}\label{si sec:fitting}
The information-space model that we propose in the main text in Eq~\ref{eq:info benefit} has four parameters that we must fit; namely, the exponent for information intake $a$, team size scaling exponent $b$, the combination of fractional team size and team scaling units $C\log(1-p)$, and the units $D$. To connect it with the data that we have on reading, we also take a scaling relation to convert the number of employees $N$ to the number of records $R$, or $N = R_0 R^\beta$. We rewrite the equation with the substitution below,
\begin{align}
	I_{\rm div} &= D R_0^a R^{a\beta} \left[ 1-e^{-C R_0^b R^{b\beta}} \right].
\end{align}
Thus, we need only fit the parameter combinations $a\beta$, $b\beta$, $CR_0^b$, and $DR_0^a$. We iwll do this for the three mappings of information variety $I_{\rm div}$ to articles, sources, and topics. For each of these cases, we must find a different set of parameters that describe how team expertise overlaps in the respective units. 

We fit the parameters by numerical optimization, but logarithmic least-squares does not ensure that the resulting functional form satisfies physical limits. There are several criteria that must be observed for the fits to be physically sensible:
\begin{enumerate}
	\item The parameters must all be positive, which also ensures that the function is positively valued in the region $R\geq1$.
	\item The function does not exceed the 1:1 line, since it is by definition impossible to read with higher variety than the number of items read.
	\item The exponent $a$ for the max must be strictly greater than or equal to the exponent $a$ for the mean; otherwise, the mean can be greater than the maximum for large $R$. 
\end{enumerate}

For the fits to articles and sources, the fits are straightforward because we have a wide range of observations, and no single firm saturates the number of unique articles or sources. We fit the function using logarithmic least-squares and implement the listed conditions using numerical constraints in NumPy. 

\begin{figure}\centering
	\includegraphics[width=.9\linewidth]{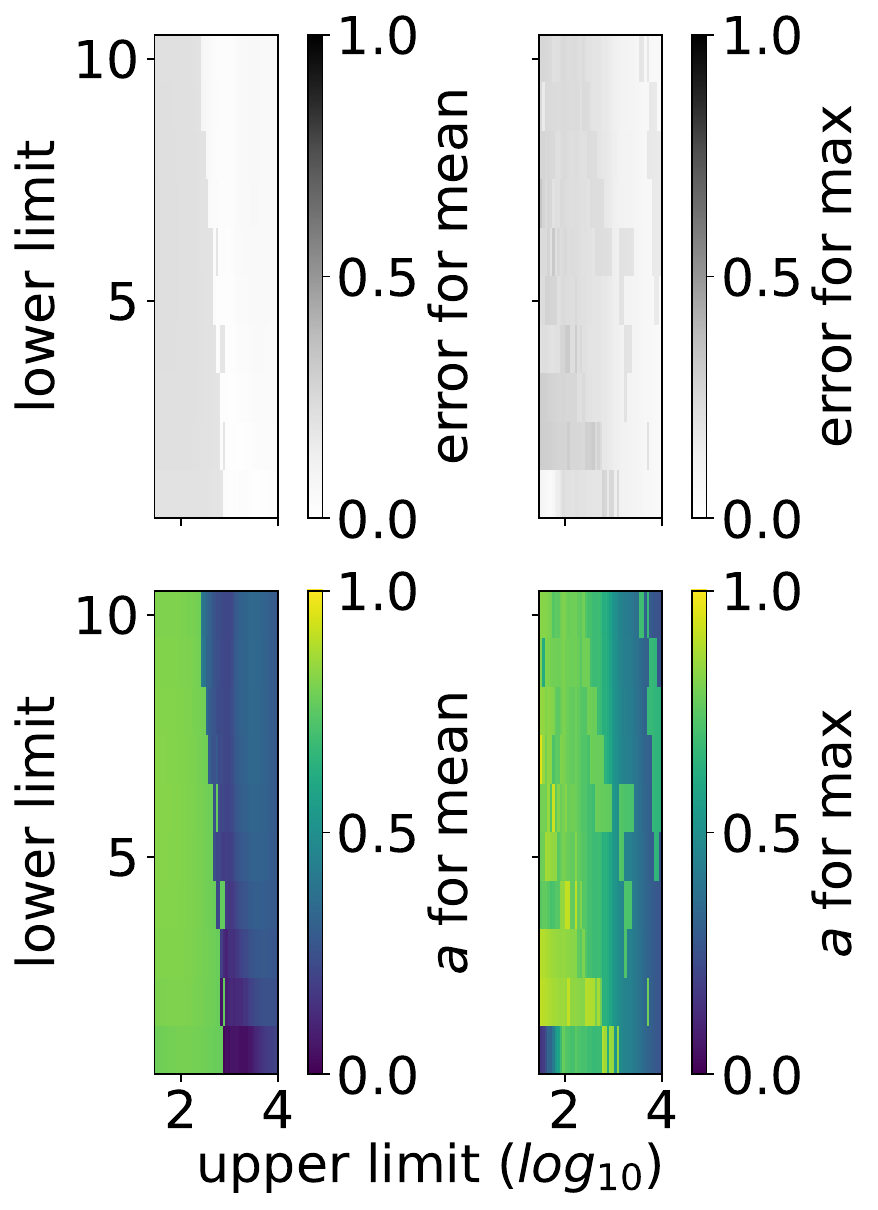}
	\caption{Variation in fit for exponent $a$ in information space model. On the left,  the normalized logarithmic errors below 1/20 and fits that have errors of about 1/5 and above.}\label{si gr:fit var}
\end{figure}

For topic scaling, we must worry about the hard, artifical cutoff in the number of topics that we observe, so it does not make sense to fit the entire curve; i.e.~we need a maximum cutoff $R_{\rm max}$. We will also have a lower cutoff because many firms that have an appearance of $R=1$ are both poorly observed and are set with an artificial lower bound: articles are labeled with typically $T=10$. To be systematic about the fit, we scan over a combination of lower and upper cutoffs and calculate the fit error as we show in Figure~\ref{si gr:fit var}c and d. The fits separate into two bands, on the left corresponding to larger $a$ and larger errors for the fit, whereas on the right side we have smaller $a$ and better fits. 

\begin{figure}\centering
	\includegraphics[width=.5\linewidth]{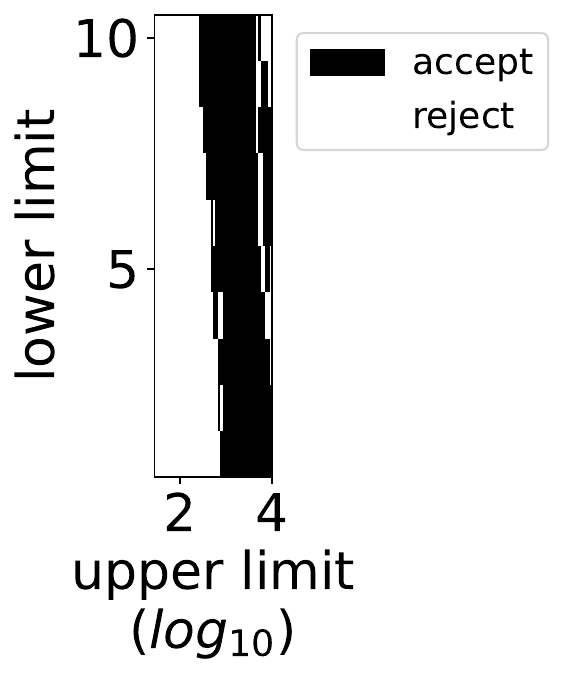}
	\caption{Region of lower and upper cutoffs for topic Heaps plot fit.}\label{si gr:choice fits}
\end{figure}

To identify ``reasonably good'' fits, we apply two criteria: first, we leverage condition 3, assuming that sufficiently informative regions of the curve will ensure compliance; we then take fits that satisfy it. The second criterion we obtain from inspecting the distribution of errors for the fit of the mean curve across the range of lower and upper cutoffs shown in Figure~\ref{si gr:fit var}. The distribution is strongly bimodal as the colors indicate, with a set of fits with normalized errors around 1/20 and below and another set with normalized errors of about 1/5 and above. Putting the two criteria together, we obtain a band of good fits that we highlight in Figure~\ref{si gr:choice fits}. These constitute the pairs of lower and upper limits that we consider for obtaining the median (best fit) and the 90\% confidence intervals that we show in the main text. As we show in Figure~\ref{gr:info strategies}, the fits hew closely to the data below the cutoff and do not overfit, deviating from it at large $T$. Most importantly, the exponent that we focus on in the main text $a$, shows a distribution that is strongly confined to $a<2/3$, which corresponds to the conservative strategy that we discuss in the main text.

\begin{table}
\caption{Scaling exponents for economic against information footprint. Error bars represent one standard deviation over bootstrapped fits.}\label{si tab:exponents}
\begin{tabular}{l|l|l|l|l}
			& assets 		& PPE 			& employees 		& sales\\
\hline
	records & $0.68\pm0.02$	& $ 0.79\pm0.02$ & $ 0.76\pm0.01$		& $ 0.77\pm0.02$ \\
	articles &$0.73\pm0.02$	& $ 0.85\pm0.02$ & $ 0.82\pm0.01$		& $ 0.82\pm0.02$ \\
	sources	& $ 0.85\pm0.02$	& $1.01\pm0.03$ & $0.97\pm0.02$		& $0.97\pm0.02$ \\
	topics &	$0.95\pm0.03$	& $0.97\pm0.03$ & $0.95\pm0.03$		& $0.95\pm0.03$ \\
\end{tabular}
\end{table}

\bibliography{refs,frefs}